%% file: main.tex
\documentclass[conference]{IEEEtran}
\IEEEoverridecommandlockouts

\usepackage{tikz}
\usetikzlibrary{automata, positioning, arrows}
\usepackage{amsmath}
\usepackage{xspace}
\usepackage{caption}
\usepackage{subcaption}
\usepackage{algorithm}
\usepackage{comment}
\usepackage[font=small,labelfont=bf, skip=5pt]{caption}
\usepackage{enumitem}
\usepackage[normalem]{ulem}
\usepackage{cleveref}
\usepackage{booktabs}
\usepackage{multirow}
\usepackage{enumitem}

\input{macros}

\usepackage{cite}
\usepackage{amsmath,amssymb,amsfonts}
\usepackage{algorithmic}
\usepackage{graphicx}
\usepackage{textcomp}
\usepackage{xcolor}
\def\BibTeX{{\rm B\kern-.05em{\sc i\kern-.025em b}\kern-.08em
    T\kern-.1667em\lower.7ex\hbox{E}\kern-.125emX}}

\begin{document}

\title{Conference Paper Title*\\
{\footnotesize \textsuperscript{*}Note: Sub-titles are not captured in Xplore and
should not be used}
\thanks{Corresponding author: Devendra Dahiphale (devphale@google.com)}
}

\include{authors_template_updated}

\title{\sysname}

\maketitle
\begin{abstract}
\input{abstract}
\end{abstract}


\input{introduction}
\input{literature}
\input{vision_and_objective}
\input{methodology_and_use_case}
\input{evaluations_and_results}

\input{generalization_guidelines}
\input{limitations}
\input{real_world_impact}
\input{conclusion}
\input{future_scope}

\bibliographystyle{plain}
\bibliography{references.bib}

\clearpage

\end{document}

%% file: macros.tex
\newcommand{\todo}[1]{{\color{red} TODO: #1}}
\newcommand{\devphale}[1]{{\color{teal} Devendra says: #1}}

\newcommand{\sysname}{\sc Enhancing Trust and Safety in Digital Payments: An LLM-Powered Approach\xspace}

%% file: authors_template_updated.tex
\author{
\IEEEauthorblockN{Devendra Dahiphale\textsuperscript{1},
Naveen Madiraju\textsuperscript{2},
Justin Lin\textsuperscript{3},
Rutvik Karve\textsuperscript{4},
Monu Agrawal\textsuperscript{5},
Anant Modwal\textsuperscript{6},\\
Ramanan Balakrishnan\textsuperscript{7},
Shanay Shah\textsuperscript{8},
Govind Kaushal\textsuperscript{9},
Priya Mandawat\textsuperscript{10},\\
Prakash Hariramani\textsuperscript{11}, and
Arif Merchant\textsuperscript{12}}

\IEEEauthorblockA{\Large\textit{Google, Inc} \\
\small\{devphale\textsuperscript{1}, naveenmadiraju\textsuperscript{2}, justlin\textsuperscript{3}, rutvikkarve\textsuperscript{4}, monu\textsuperscript{5}, anantmodwal\textsuperscript{6},\\ ramananb\textsuperscript{7}, shanayshah\textsuperscript{8}, gkaushal\textsuperscript{9}, priyamandawat\textsuperscript{10},
phariramani\textsuperscript{11},
aamerchant\textsuperscript{12}\}@google.com}
}

%% file: abstract.tex
Digital payment systems have revolutionized financial transactions, offering unparalleled convenience and accessibility to users worldwide. However, the increasing popularity of these platforms has also attracted malicious actors seeking to exploit their vulnerabilities for financial gain. To address this challenge, robust and adaptable scam detection mechanisms are crucial for maintaining the trust and safety of digital payment ecosystems. This paper presents a comprehensive approach to scam detection, focusing on the Unified Payments Interface (UPI) in India, Google Pay (GPay) as a specific use case. The approach leverages Large Language Models (LLMs) to enhance scam classification accuracy and designs a digital assistant to aid human reviewers in identifying and mitigating fraudulent activities. The results demonstrate the potential of LLMs in augmenting existing machine learning models and improving the efficiency, accuracy, quality, and consistency of scam reviews, ultimately contributing to a safer and more secure digital payment landscape. Our evaluation of the Gemini Ultra model on curated transaction data showed a 93.33\% accuracy in scam classification.  Furthermore, the model demonstrated 89\% accuracy in generating reasoning for these classifications. A promising fact, the model identified 32\% new accurate reasons for suspected scams that human reviewers had not included in the review notes.

%% file: keywords.tex
Trust and Safety, Digital payment systems, Large Language Models, Deep Learning, Payments Security, UPI, Fraud Detection, Reasoning.

%% file: introduction.tex
\section{Introduction} \label{sec:introduction}
Digital payment systems have become an integral part of modern life, facilitating seamless and secure transactions across various platforms. The convenience and accessibility of these systems have democratized access to financial services and driven a surge in digital transactions. However, the widespread adoption of digital payments has also attracted scammers and fraudulent actors seeking to exploit their vulnerabilities for financial gain. The financial losses incurred due to these fraudulent activities can be substantial, not only for individual users but also for the payment platforms themselves, with some estimates suggesting that annual losses due to scams in India alone could reach billions of dollars \cite{gupta2024online, kapron2023google}.

To maintain the trust and safety of digital payment ecosystems, robust and adaptable scam detection mechanisms are imperative. Traditional machine learning (ML) \cite{jordan2015machine_learning_foundational} models, while effective to some extent, often struggle to keep pace with the evolving tactics of scammers. These models typically rely on structured data and predefined rules, which may not be sufficient to capture the subtle nuances and contextual cues that are often present in scam transactions. Moreover, the sheer volume of transactions processed by digital payment platforms can overwhelm traditional ML models, leading to delays in detection and potential financial losses.

In addition to ML models that can automatically detect and block scam where there is a high degree of confidence in detection, many transactions also require a "human review" where the signals are less clear / more ambiguous. As such, a team of human reviewers trained on scam detection acts as a fallback mechanism to detect and prevent scam that is not easily caught by ML models.

Therefore, both scam classification and the reasoning behind those classifications offer promising use cases for Large Language Models (LLMs) \cite{llm_survey_minaee2024large}.  LLMs can act as classifiers to directly detect and block scams. Additionally, when trained on expert knowledge, they can serve as digital assistants for human reviewers, empowering them to make faster, consistent, more informed decisions by identifying key indicators, interpreting them accurately, and predicting potential outcomes (scam or not scam).

The integration of LLMs into the scam detection workflow is motivated by their ability to process and interpret textual data, which is often crucial in identifying fraudulent activities. LLMs can analyze transaction descriptions, user messages, and other textual cues to discern patterns that may not be readily apparent to traditional ML models. Furthermore, LLMs can be trained on large datasets of labeled scam and non-scam transactions, enabling them to learn the subtle differences between legitimate and fraudulent activities. Despite the potential of LLMs, their application in scam detection within digital payment systems, particularly in the context of Unified Payments Interface (UPI) transactions in India, remains largely unexplored.

In the context of UPI \cite{npci_website} transactions in India, the problem of scam detection is particularly acute due to the sheer volume and diversity of transactions. In India, UPI transactions account for \$1.7 trillion annually, and GPay processes about a third of all UPI transactions \cite{kapron2023google}. Google Pay (GPay), as a leading UPI-enabled payment platform, processes a massive number of transactions daily, making it a prime target for scammers \cite{signs_of_fraud_Li_2012}. The development of an LLM-based scam detection solution for GPay can serve as a valuable case study for addressing the broader challenge of scam detection in UPI transactions and other digital payment systems.

This paper presents a comprehensive approach to scam detection in digital payments, focusing on UPI transactions in India and GPay as a specific use case. While this research outlines results from UPI and GPay as the tested use case, we believe the method is broadly applicable to other payment systems globally and also to scam classification and digital assistants beyond these areas. The approach leverages LLMs to enhance scam classification accuracy and introduces a digital assistant to aid human reviewers in identifying and mitigating fraudulent activities. The research encompasses data preprocessing, model training, evaluation, and the development of a reasoning engine to provide human-interpretable explanations for scam classifications.

Evaluation of the Gemini Ultra model (an earlier released version) on curated transaction data showed a 93.33\% accuracy in scam classification. Furthermore, the model demonstrated 89\% accuracy in generating reasoning for these classifications. It also identified 32\% new accurate reasons for suspected scams that human reviewers had overlooked.  The results showed high recall (above 90\% consistently), which is most important in scam classification. The results demonstrate the potential of LLMs in augmenting existing machine learning models and improving the efficiency and accuracy of scam reviews, ultimately contributing to a safer and more secure digital payment landscape.

%% file: literature.tex
\section{Literature Survey} \label{sec:literature_survey}
The application of Machine Learning (ML) \cite{anomaly_detection_in_financial_data_using_ml_and_llm} in financial fraud detection has been a topic of growing interest in recent years. Numerous studies have explored various ML algorithms, such as decision trees \cite{cc_fraud_detection}, support vector machines \cite{Sahin2011DetectingCC}, and neural networks \cite{deep_learning_fraud_detection}, for detecting fraudulent transactions in credit cards \cite{credit_card_fraud_detection_using_ml_sl}, online payments \cite{fraud_signs_using_data_mining}, and insurance claims. Studies have shown the effectiveness of LLMs in various security contexts, including credit card fraud detection , anomaly detection in financial data \cite{anomaly_detection_in_financial_data_using_ml_and_llm, jiang2024detectingscamsusinglarge}, and ensuring data protection in e-commerce transactions \cite{amincommerce_llm}. These models have demonstrated proficiency in analyzing large datasets, identifying patterns indicative of fraudulent activities, and responding to potential threats in real-time.

In addition to traditional ML approaches, researchers have explored the potential of deep learning for fraud detection \cite{financial_fraud_detection_model_using_dl_lstm, credit_card_fraud_detection_using_ml_sl}. Deep learning models, such as autoencoders and recurrent neural networks (RNNs) \cite{fraud_detection_rnn,credit_card_fraud_detection_using_ml_sl} have shown promising results in capturing complex patterns and anomalies in financial data. Some studies have also investigated the use of LLMs for scam detection in specific contexts, such as email phishing scams \cite{phishing_detection_using_nlp_and_ml}.  However, their application in the context of digital payment systems, particularly UPI transactions in India, remains a nascent area of research.

This paper aims to bridge this gap by exploring the potential of LLMs in detecting scams in UPI transactions, focusing on Google Pay (GPay) as a use case. The paper builds upon existing research in ML-based fraud detection and leverages the unique capabilities of LLMs to analyze textual data and identify subtle patterns indicative of fraudulent activities. Notably, our approach draws inspiration from recent advancements in LLMs for tabular data classification, such as TabLLM \cite{hegselmann2023tabllm} and TabText \cite{carballo2023tabtext}, which have demonstrated promising results in few-shot learning scenarios.

%% file: vision_and_objective.tex
\section{Vision and Objective} \label{sec:vision_and_objective}
In this paper we develop a comprehensive and adaptable scam detection solution using LLMs for digital payment systems, with a specific focus on UPI transactions in India, Google Pay (GPay) as a primary use case. We aim to achieve the following objectives:

\begin{itemize}
    \item \textbf{LLM Classifier (Enhance Scam Classification Accuracy):} Leverage LLMs to improve the accuracy of scam classification by analyzing tabular data (both numerical and textual) and identifying subtle patterns indicative of fraudulent activities. Reduce the reviewing burden on human reviewers by automating the identification of potential scams. \\
    To evaluate the usefulness of LLMs in classifying scam transactions we focus on assessing the performance of Gemini \cite{geminiteam2024geminifamilyhighlycapable}, an LLM, in accurately identifying scam transactions. Using different prompting techniques and fine-tuning on GPay India data, the goal is to achieve a specified level of precision, recall, and confidence in scam detection to demonstrate the significant potential of LLMs for scam detection.
    
    \item \textbf{LLM Digital Assistant with Reasoning:} Alternatively, we refer Digital Assistant by Reasoning Engine throughout the paper. Create a digital assistant that can assist human reviewers by providing reasoning and explanations for scam classifications, thereby improving the efficiency and accuracy of manual reviews. The Digital Assistant aims to streamline the review process by highlighting key features of interest to scam reviewers quickly and providing insights into the rationale behind scam classifications. \\
    To achieve this, we design and build an LLM-based Reasoning Engine and Digital Assistant (that also has classification capabilities) that can reason about transaction features to detect potential scams from provided guidelines. The Digital Assistant provides human interpretable explanations for scam or non-scam classifications, assisting human reviewers in making informed decisions and evaluating its accuracy. The digital assistant synthesizes signals and features from transactions and accounts, presenting them in an easily understandable format.
    
    \item \textbf{Generalizable Solution for scam detection:} Propose a scam detection approach with reasoning that is not only effective for UPI and payments transactions but can also be generalized and applied to other digital payment systems. This involves creating a flexible and adaptable approach that can be easily applied to different payment platforms and transaction types.
    
\end{itemize}

%% file: methodology_and_use_case.tex
\begin{figure*}[ht!]
    \centering
    \includegraphics[scale=0.5]{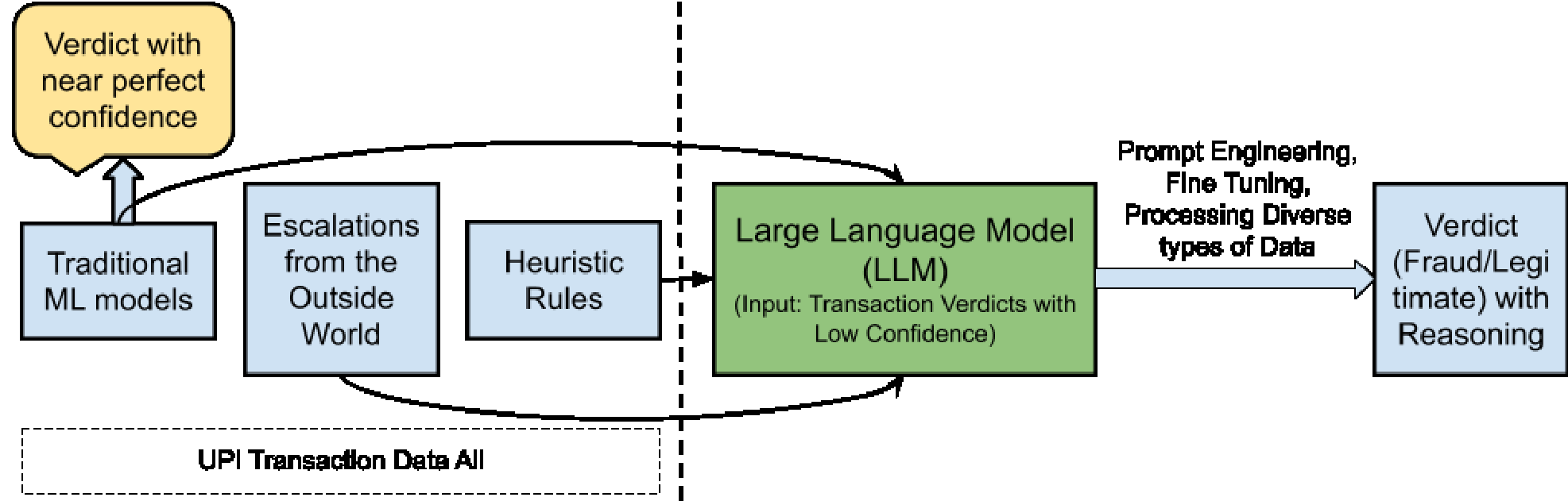}
    \caption{Input data for LLM Model (Gemini)}
    \label{fig:llm_input}
\end{figure*}

\section{Methodology \& Use Case} \label{sec:methodology_and_use_case}

In this section we detail the implementation of Large Language Models (LLMs) for the purpose of scam classification. The approach involves two key components: the development of an LLM Classifier and a Reasoning Engine. We present an LLM classifier and a reasoning engine using Google's Gemini to improve scam detection in GPay India and augment current ML models that are used for scam detection. Currently, GPay transactions undergo various checks to distinguish between legitimate and fraudulent activities and existing models are able to detect some scams with very high confidence and others with a lower but still significant confidence. Scams that are detected with high confidence are Auto-Denied while others get sent for manual review. Our work improves the latter transaction classification by augmenting current classifiers with LLM models that provide extra reasoning for each verdict.
\subsection{Data Sources and Preparation} \label{sec:data_sources_and_preparation}
Figure \ref{fig:llm_input} elucidates the input data utilized in the study.
The LLM classifier was trained on datasets containing up to 4 million transactions, while the reasoning engine was run using a curated set of approximately 40 transactions. The experiments involved varying the number of transactions used for training to assess its impact on model performance. \\
The features encompass information about the order itself (such as amount, type, memo, and order text), payer \& payee behavior (including transaction history and payment methods), and external factors like spam reports. We applied the following techniques to prepare our dataset:

\noindent \textbf{Feature Selection \& Anonymization}: We carefully selected a subset of features (top 20) encompassing transaction details, payer and payee information, and their relationships. To protect user privacy, all data was anonymized prior.

\noindent \textbf{Data Transformation}: The numerical features were converted into a text-friendly format that the LLM can interpret. To bridge the gap between numerical data and the text-based nature of LLMs, we applied several transformations:
\begin{itemize}
    \item \textbf{Serialization}: We transformed numerical values such as transaction amounts into descriptive categories (e.g., "very low," "low," "medium," "high" "very high") based on quantiles calculated from the training data \cite{wang2023anypredict} \cite{carballo2023tabtext}.
    \item \textbf{Missing Value Handling:} We addressed missing values by replacing them with a placeholder like "unknown" due to its simplicity and interpretability by the LLM.
    \item \textbf{Data Splitting \& Balancing}: To ensure robust model evaluation and mitigate potential bias from class imbalance, we partitioned the data into training, validation, and test sets using a stratified split strategy. This ensured a balanced representation of both scam and non-scam transactions across all sets.
\end{itemize}

\subsection{Models Used}
Our work utilized various Gemini models, leveraging their unique strengths: \\
\textbf{Gemini Flash}: Employed initially for efficient scam classification. \\
\textbf{Gemini Ultra}: Used for advanced scam classification incorporating reasoning capabilities. \\
\textbf{Gemini Pro and Nano}: Explored for learning and development purposes. \\
Our primary findings are presented in Section  uses Gemini Flash's and Ultra's earlier versions.

Our objective is to demonstrate the broader strategic advantages and future possibilities of LLMs in combating fraud and improving security within the digital payment landscape and also in general trust \& safety domain.

\begin{figure*}[ht]
    \centering
    \includegraphics[scale=0.48]{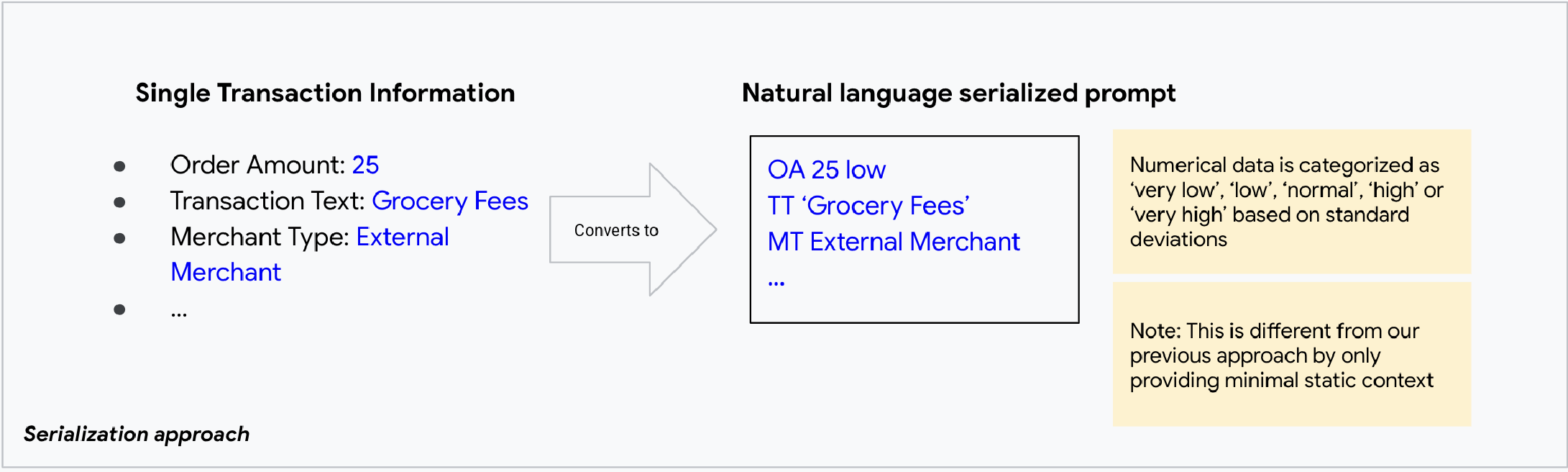}
    \caption{Data Transformation Example}
    \label{fig:approach_in_brief}
\end{figure*} \label{sec:use_case}

\begin{figure*}[ht]
    \centering
    \includegraphics[scale=0.55]{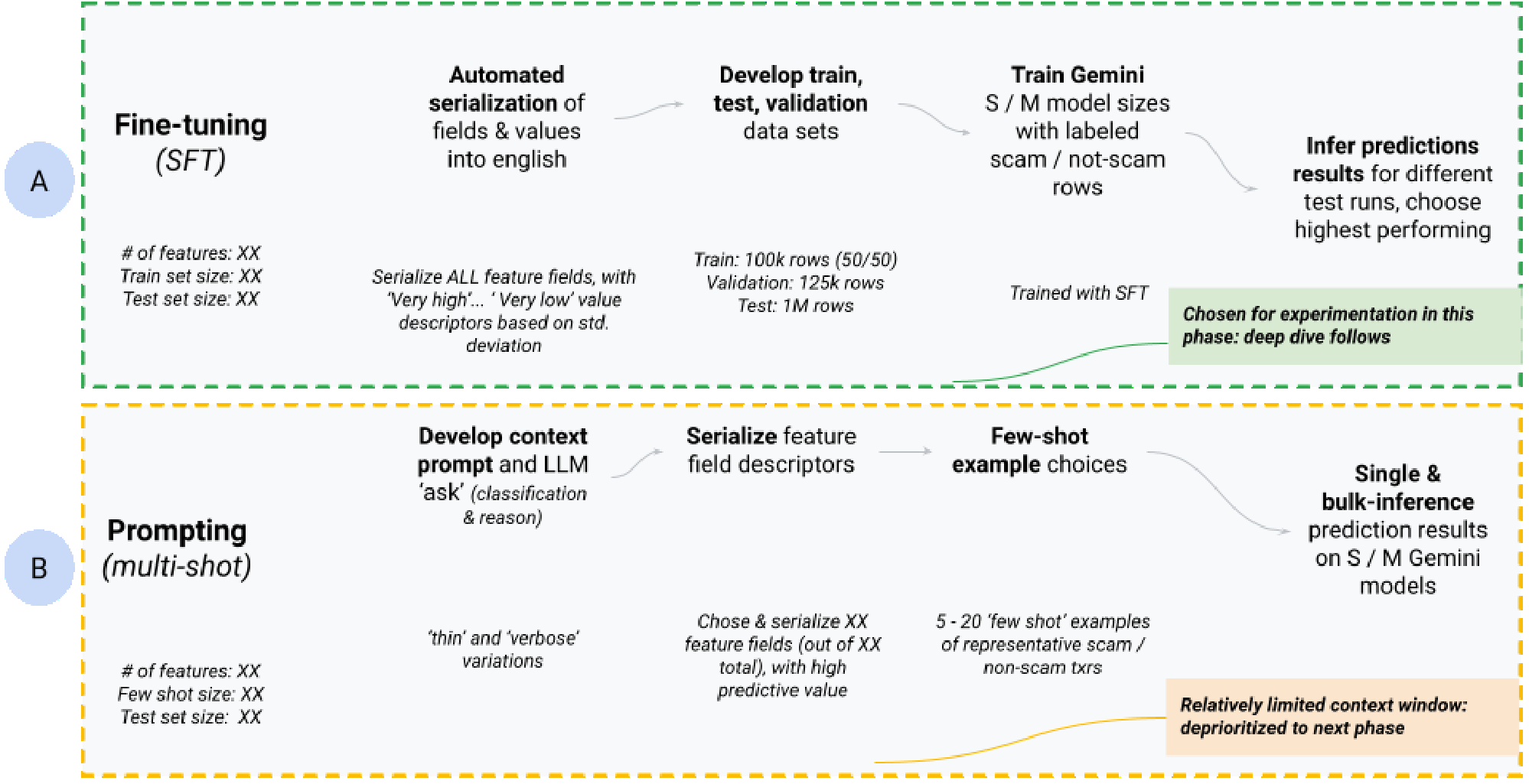}
    \caption{LLM Fine-Tuning}
    \label{fig:llm_fine_tuning_approach}
\end{figure*} \label{sec:llm_fine_tuning_approach}

\begin{figure*}[ht]
    \centering
    \includegraphics[scale=0.65]{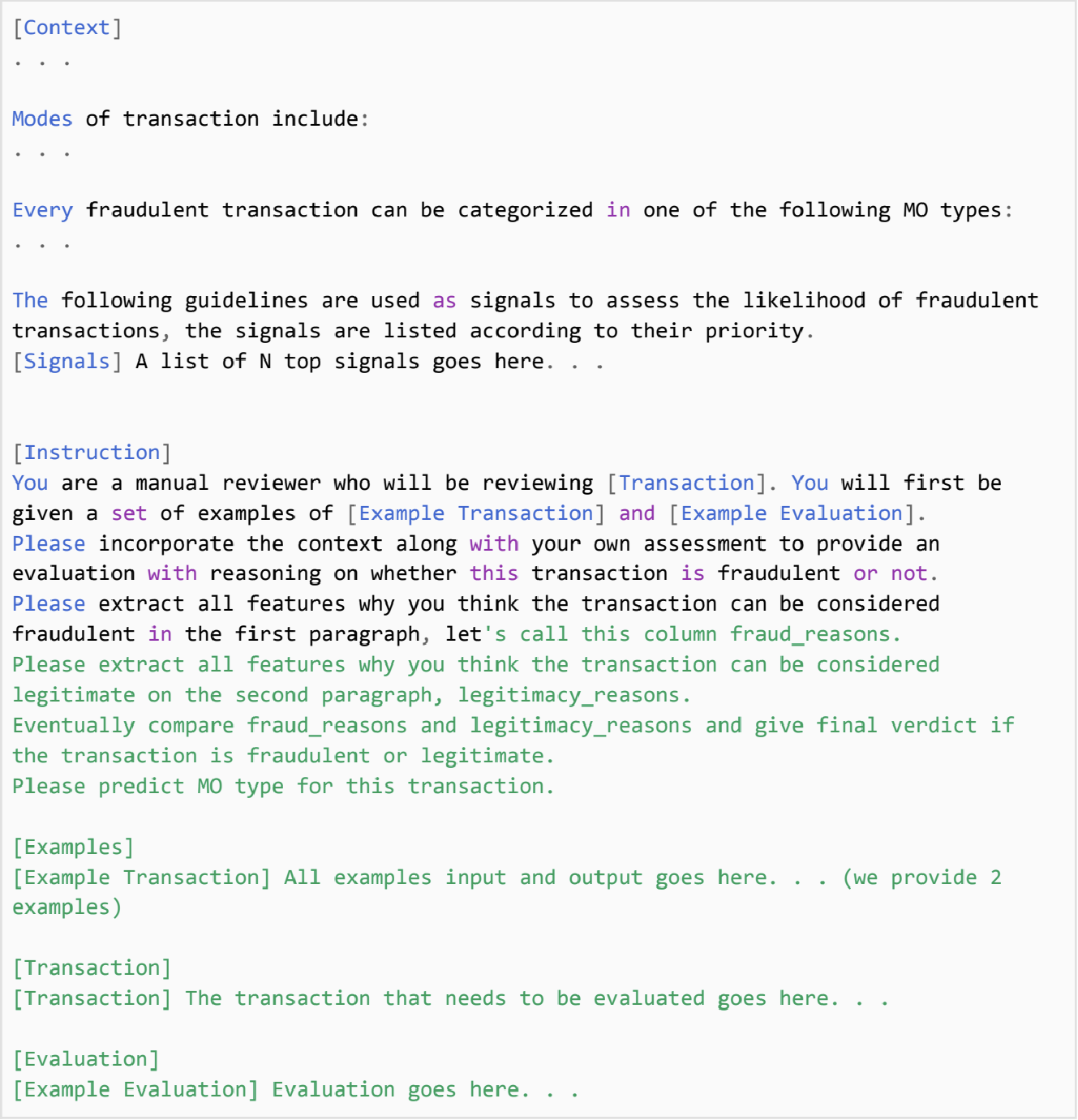}
    \caption{LLM Classifier with Reasoning: A Few-Shot Prompting Skeleton}
    \label{fig:llm_classifier_with_reasoning_few_shot_prompting_example}
\end{figure*}

\subsection{LLM Classifier}
The LLM Classifier, on top of feature engineering, data cleaning and preprocessing, employs few-shot prompting and fine-tuning. For both of these strategies prompts are used to train the model. The primary goal of the classifier prompts are to guide the LLM in determining whether a given UPI transaction is fraudulent or legitimate. A classifier prompt is a set of instructions that guide an LLM to classify a given input. To achieve this, the prompt includes:
\begin{itemize}
    \item \textit{Context}: Provides background information about GPay, the UPI system, and common scam types. This helps the LLM understand the specific domain it's working in;
    \item \textit{Feature Descriptions}: Explains the different features present in the transaction data, such as transaction amount, merchant type, order type, etc. This ensures the LLM knows what information is available to it;
    \item \textit{Instructions}: Clearly states the task: classify the transaction as "scam" or "not scam" and asks for a brief explanation. This sets the expectation for the LLM's output;
    \item \textit{Examples}: Some prompts may include a few examples of labeled transactions and their explanations. This can help the LLM learn the patterns to look for. 
\end{itemize}

These classifier prompts were then used as follows:

\noindent \textbf{Few-shot prompting:} Involves providing the LLM with a limited number of examples of scam and non-scam transactions, along with explanations for the classifications. This approach leverages the LLM's ability to learn from a few examples and generalize to new, unseen instances.

\noindent \textbf{Fine-tuning:} Involves training the model on a larger labeled dataset to optimize its performance on the specific task of scam detection. Prompt design is critical also for fine-tuning, with prompts carefully crafted to provide context and instructions. The model is trained to classify transactions as "scam" or "not scam". Prompts include contextual information about GPay, common scam types, and features used for classification.

In addition to prompting and fine-tuning, model variations and experiments were also employed to train the LLM classifier. During Phase 1, various experiments were conducted to assess the impact of different model sizes and training datasets on the LLM's performance. Gemini Flash, Gemini Pro, Gemini Ultra,  and Gemini Nano models were trained on datasets ranging from 100K to 1 million transactions. The experiments revealed that larger fine-tuning datasets significantly boosted the model's performance, increasing both precision and recall. Additionally, it was observed that smaller models like Gemini Nano could achieve comparable performance to larger models with minimal loss, suggesting the potential for more efficient and cost-effective deployment in production environments.

\subsection{Reasoning Engine and Digital Assistant}
Building on the LLM classifier from Phase 1, Phase 2 focused on developing a digital assistant to aid human reviewers in the scam review process. This assistant provides the top N reasons why a transaction could be a scam and the top N reasons why it might be legitimate. Along with these reasons "for" and "against," the LLM provides a verdict on whether the transaction is fraudulent, improving the accuracy, efficiency, and consistency of manual reviews.

Central to this digital assistant is the reasoning engine. This engine generates human-interpretable explanations for scam classifications by analyzing transaction data and identifying the most salient features contributing to the LLM's verdict. It then synthesizes these features into a coherent explanation, highlighting the specific aspects of the transaction that raised suspicion. To achieve this, the reasoning engine was trained on a curated dataset of scam and non-scam transactions, including detailed reviewer notes explaining the rationale behind each classification. This training allows the engine to learn intricate patterns and features associated with different types of scams and generate explanations aligned with human reviewers' reasoning processes.

The explanations are designed for easy comprehension by human reviewers, ensuring they can quickly grasp the rationale behind a scam classification and make informed decisions. These explanations are not mere restatements of the data but rather a synthesis of the most relevant information, presented clearly and concisely. The reasoning engine employs a structured approach mirroring a human reviewer's thought process. It identifies key features indicative of a scam, such as unusual transaction patterns, suspicious merchant behavior, or anomalies in user profiles. The engine then analyzes these features' values, comparing them to established norms and thresholds, and synthesizes this information into a comprehensive explanation highlighting why a transaction was flagged as suspicious. 

To build the Reasoning Engine, the LLM is prompted on a curated dataset of transactions, accompanied by detailed reviewer notes that explain the reasoning behind each classification. Prompt engineering played a crucial role in refining the reasoning engine's capabilities. The prompts were designed to elicit specific types of responses from the LLM, such as identifying the most important features, explaining their significance, and suggesting potential modus operandi (MO). The prompts were iteratively refined based on feedback from human reviewers to ensure that the explanations were accurate, relevant, and helpful. The final version of the prompt is as shown in Figure \ref{fig:llm_classifier_with_reasoning_few_shot_prompting_example}, which includes the following components:

\noindent \textbf{Context:}
    \begin{itemize}
        \item Modes of transaction: Provides information on the different ways a transaction can be initiated, like payer\_initiated\_lookup, app\_intent, qr\_scan, or payment\_request.
        \item MO types: Outlines the categories of fraudulent activities (e.g., impersonation, phishing) to help the reasoning engine identify potential scams.
        \item Signals: Includes a list of the top N signals used to assess the likelihood of fraud. These could be factors like transaction amount, transaction text, user history, merchant type, etc. One of the most important things to note here is that rearranging the order of these signals changes the final results. We found out that ordering these signals according to their priority yields the best results. We also observed that converting numerical values into textual formats with corresponding explanations led to a substantial improvement in results.
    \end{itemize}

\noindent \textbf{Instruction:}
    \begin{itemize}
        \item Role: The reasoning engine is instructed to act as a manual reviewer, implying a need for careful scrutiny and logical reasoning.
        \item Examples: The reasoning engine is provided with a set of example transactions and their corresponding evaluations. This helps to learn the patterns and reasoning used in scam identification.
        \item Task: The reasoning engine must evaluate the transaction and determine if it's fraudulent. It needs to extract features supporting fraud and legitimacy separately.
        \item Reasoning: The reasoning engine is asked to compare the fraud and legitimacy reasons, reach a final verdict (fraudulent/legitimate), and predict the MO type if it's a scam.
    \end{itemize}

\noindent \textbf{Examples:} This section includes 2 example transactions with their evaluations. The examples serve as a guide for the reasoning engine to understand how to analyze transactions and provide structured explanations.
    
\noindent \textbf{Transaction:} The actual transaction data that the reasoning engine needs to evaluate is provided here. It would likely include details like the transaction amount, merchant, user information, transaction description, and other relevant features.

\noindent \textbf{Evaluation:} This is where Robin provides its assessment. It should start with features supporting fraud, then features supporting legitimacy. It should conclude with a clear verdict and, if applicable, the predicted MO type.

The example of a prompt used to elicit reasoning from the LLM is shown in Figure \ref{fig:llm_classifier_with_reasoning_few_shot_prompting_example}. This prompt, when used with a powerful LLM like Gemini, significantly enhances scam detection. The reasoning engine's ability to analyze textual data, identify patterns, and employ structured reasoning leads to more accurate and efficient identification of fraudulent transactions.  Furthermore, the generated explanations provide invaluable insights to human reviewers, enabling them to understand why a transaction was flagged and make more informed decisions.

%% file: evaluations_and_results.tex
\section{Evaluations and Results} \label{sec:evaluation_and_results}
\subsection{LLM as a Classifier Results}
\begin{figure*}[htp]
    \centering
    \includegraphics[scale=0.4]{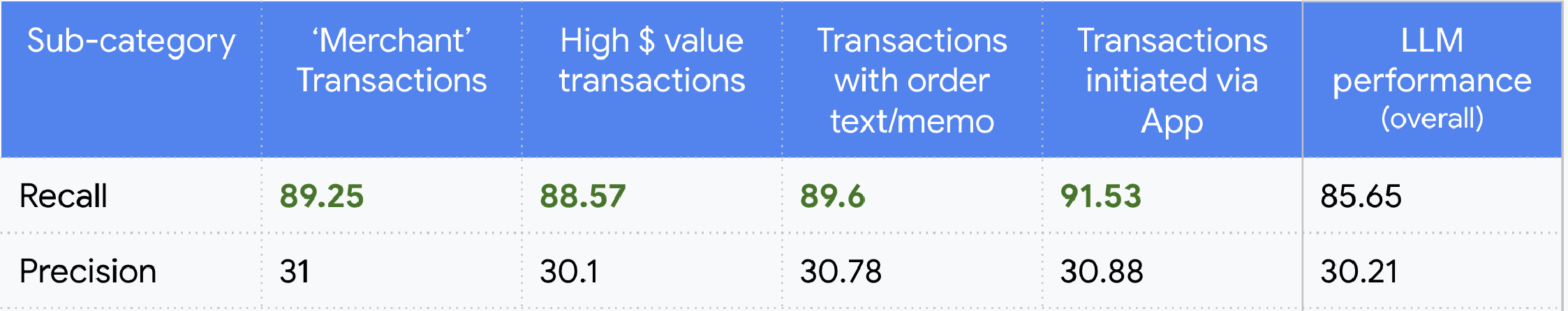}
    \caption{LLM Classifier Overall Performance \& Sub-segments}
    \label{fig:llm_classifier_results_and_subsegments}
\end{figure*}

The LLM classifier's performance is evaluated using a held-out test set of labeled UPI transactions. The evaluation metrics included precision, recall, F1 score, and area under the receiver operating characteristic curve (AUC-ROC). The results (as shown in Figure \ref{fig:llm_classifier_results_and_subsegments}) demonstrated the LLM classifier's effectiveness in identifying scam transactions, achieving a high recall rate while maintaining reasonable precision. The model's ability to analyze textual data, such as transaction descriptions, proved to be a significant advantage, enabling it to capture subtle patterns indicative of fraudulent activities.

In addition to overall performance, the LLM's performance was assessed on specific transaction segments, such as external merchant transactions, high-value transactions, and transactions initiated via applications. The LLM exhibited varying performance across these segments, highlighting the need for further refinement and customization to address the unique characteristics of different transaction types.

\begin{table*}[ht]
\centering
\caption{Summary of LLM Fine-tuning Experiments for Scam Detection}
\label{tab:classifier_learnings}
\begin{tabular}{|p{0.2\linewidth}|p{0.35\linewidth}|p{0.38\linewidth}|}
\hline
\textbf{Experiment} & \textbf{Description} & \textbf{Key Findings/Conclusions} \\
\hline
\hline
Usefulness of Numeric Data & Tested combinations of prompts including and excluding raw and categorical numeric data. & Raw numeric data is crucial for model performance.  Further investigation is needed to determine the utility of categorical numeric data.  \\
\hline
 Data from two different periods & Trained models on data from different time periods (April-May vs. July-August) and evaluated on November data. & Models exhibited similar performance to the baseline (V1N), with higher recall observed at a 0.9 confidence threshold. This suggests consistent model performance and accuracy for input data across different time-frames. \\ 
\hline
Gemini Nano Model  & Trained the smaller Gemini Nano model instead of Gemini Flash. & Observed a slight performance regression compared to Gemini Flash with 500 training steps. However, increasing training steps to 1000 yielded comparable results to Gemini Flash, indicating the potential for efficient deployment.  \\
\hline
Removing Textual Context  & Removed most textual context from the prompts, except for the order memo text, to assess the impact of contextual information.  &  Model performance (precision and recall) remained largely unchanged. This indicates that textual context did not significantly contribute to the model's performance with our current prompting approach. \\ 
\hline
Increasing Training Data  &  Experimentally increased the training dataset size from 100k to 1M and then to 4M examples. &  Increasing to 1M examples resulted in significant gains in precision and recall. However, the increase to 4M examples did not yield a substantial further improvement. This suggests that there is a point of diminishing returns with training data volume.  \\
\hline
Model Convergence Analysis &  Monitored model convergence during training to determine the optimal number of training steps. &  LLM models converged towards a stable performance after a specific number of training steps, suggesting that training efficiency can be optimized by carefully monitoring convergence.  \\
\hline
Performance on Specific Transaction Segments &  Evaluated model performance on distinct transaction segments (like external merchant transactions, high-value transactions, and app-intent initiated transactions) to analyze model adaptability. & The model exhibited varying performance across these segments, highlighting the need for further refinement and customization to address the unique characteristics of different transaction types.  \\
\hline
\end{tabular}
\end{table*}

\begin{itemize}
    \item Learning \& Discussions: Table \ref{tab:classifier_learnings} summarizes the outcomes of various experiments conducted to fine-tune the LLM for scam detection. The experiments explored the impact of different factors, such as the inclusion of numeric data, the time period of the training data, the size of the LLM model, the presence of textual context, the volume of training data, and the assessment of model convergence. The table highlights key findings and conclusions drawn from each experiment, offering insights into the optimal configuration and strategies for effective LLM-based scam detection. Overall, it suggests that while larger models and datasets can improve performance, smaller models can still achieve comparable results, making them a viable option for real-world deployment.
    \item Overall Performance \& Sub-segments:The LLM classifier demonstrated promising performance across various transaction sub-categories as shown in Figure \ref{fig:llm_classifier_results_and_subsegments}. Notably, it achieved a high recall rate, exceeding 88\% in all evaluated categories, indicating its effectiveness in identifying relevant transactions. The classifier particularly excelled in identifying transactions initiated via applications, achieving a recall of 91.53\%.
    
    While the precision scores were consistently around 30\%, indicating room for improvement in reducing false positives, the model showed a slight advantage in handling transactions with order text/memo and those initiated through the applications. Overall, the LLM classifier holds significant potential for transaction classification tasks, particularly in scenarios where high recall is prioritized. These findings highlight the strengths and areas for further refinement of the LLM classifier. Future work could focus on improving precision, potentially through techniques like active learning or incorporating additional features.
    \item Observation on High Recall \& Low Precision: In the context of fraud detection, the primary goal is to identify and prevent as many fraudulent transactions as possible. This makes recall, or the ability to correctly identify true positives (actual fraud cases), the most critical metric. Even if precision is low, meaning there are many false positives (legitimate transactions flagged as fraud), it's generally less problematic because there are additional checks including final reviews by human experts.
    
    The cost of missing a fraudulent transaction (a false negative) is typically much higher than the cost of inconveniencing a user with an additional review (a false positive). Therefore, a system that prioritizes high recall, even at the expense of lower precision, is often preferred in fraud detection scenarios. The false positives can be filtered out in subsequent stages, ensuring that legitimate users are not unduly affected while minimizing the financial and reputational damage caused by undetected fraud.
\end{itemize}

\subsection{LLM as a Reasoning Engine Results}
The reasoning engine's performance was evaluated through a series of user studies involving experienced scam reviewers. The reviewers were presented with a set of scam and non-scam transactions, along with the LLM's classifications and explanations generated by the reasoning engine. The reviewers were asked to assess the accuracy, relevance, and helpfulness of the explanations.

\begin{figure}[htp]
    \centering
    \includegraphics[scale=0.5]{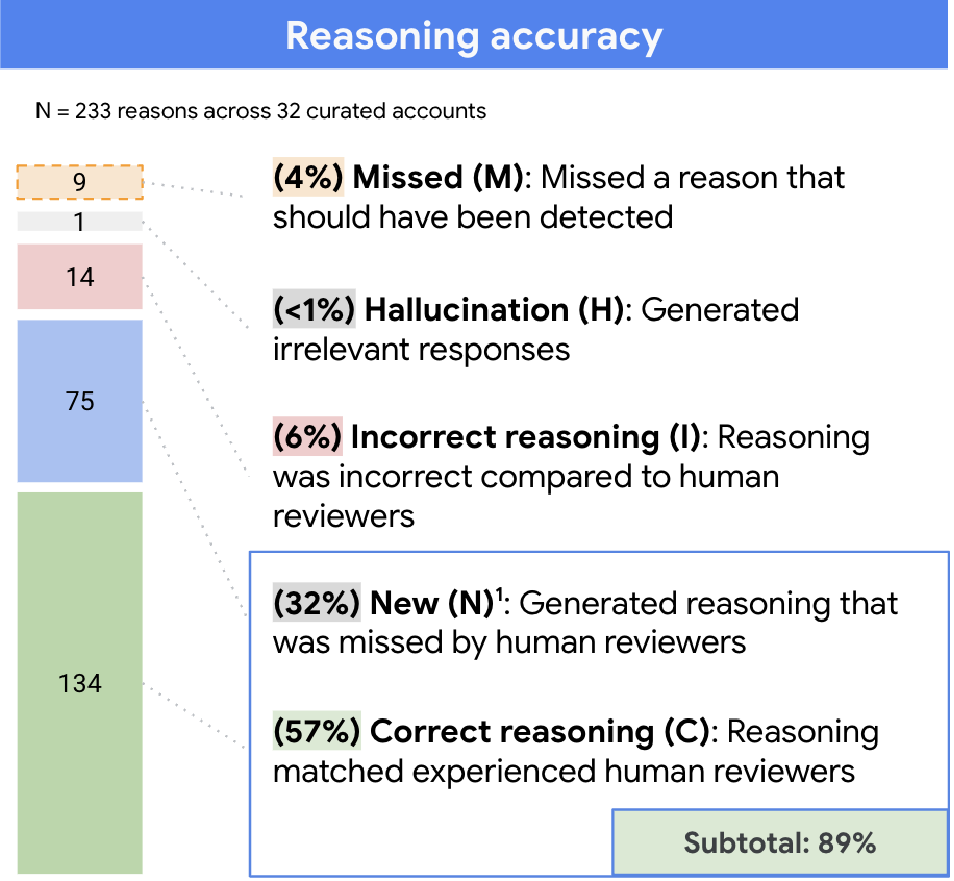}
    \caption{Reasoning Accuracy}
    \label{fig:reasoning_accuracy}
\end{figure}

\begin{figure}[htp]
    \centering
    \includegraphics[scale=0.5]{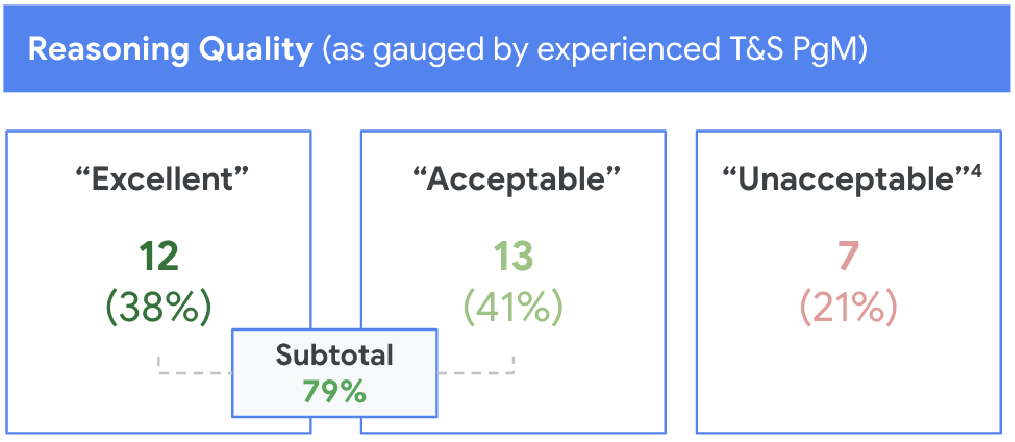}
    \caption{Reasoning Quality}
    \label{fig:reasoning_quality}
\end{figure}

The accuracy of scam classification within the reasoning engine was 93.33\%, whereas Figure \ref{fig:reasoning_accuracy} shows the accuracy of generated reasoning for those scam classifications and Figure \ref{fig:reasoning_quality} shows quality of generated reasoning. The evaluation focused on a smaller set of high-quality account reviews (40 transactions, consisting more than 300 scam reasons) to ensure a rigorous assessment of digital assistant's capabilities, comparing its outputs to high-quality human reviewer notes. The key findings are as given below.
\begin{itemize}
    \item Reasoning Accuracy: A key aspect of the evaluation focused on Reasoning Engine's ability to generate accurate reasoning for scam classifications. By comparing Reasoning Engine's outputs to a curated set of high-quality human reviewer notes, we found that Reasoning Engine achieved an 89\% accuracy rate in providing reasoning that either matched human reviewers or offered new, valid insights. This high accuracy rate suggests that Reasoning Engine can effectively replicate and even enhance human decision-making processes in scam detection.
    \begin{itemize}
        \item Correct Reasoning (C): In 57\% of the cases, Reasoning Engine's reasoning aligned perfectly with that of experienced human reviewers, demonstrating its ability to learn and apply established scam detection patterns.
        \item Incorrect Reasoning (I): A small percentage (6\%) of Reasoning Engine's reasoning was deemed incorrect compared to human reviewers, indicating areas for further refinement and learning.
        \item Hallucination (H): Less than 1\% of Reasoning Engine's responses were irrelevant or nonsensical, highlighting the need for ongoing monitoring and improvement of the model's output generation.
        \item Missed (M): In 4\% of cases, Reasoning Engine failed to identify a reason that human reviewers had detected, suggesting potential gaps in the model's understanding or application of certain scam indicators.
        \item New (N): Notably, Reasoning Engine generated new reasoning in 32\% of cases that human reviewers had not added in reviewers notes, showcasing its potential to uncover novel patterns and insights that could enhance scam detection efforts in case human reviewers missed those. The additional 32\% of reasons did not improve the accuracy of classification within the chosen accounts. However, they did reduce the time reviewers spent on reviews.
    \end{itemize}
    \item Reasoning Quality: To assess the quality of Reasoning Engine's reasoning, experienced Trust and Safety Program Managers (PgMs) evaluated its outputs. The results were promising, with 38\% of reasons rated as "Excellent" and 41\% as "Acceptable." This positive feedback indicates that Reasoning Engine's reasoning is not only accurate but also well-structured, coherent, and valuable for human reviewers, totaling 79\% positive feedback.
    \item Additional Considerations: The evaluation also considered factors such as the limited number of data fields used (12 out of all possible fields) and the manual nature of the evaluation process. While these factors may have influenced the results, the overall findings strongly suggest that Reasoning Engine has the potential to significantly improve the efficiency and accuracy of scam reviews.
    \item Extensibility and Scalability: Our experiments in other projects show that the approach used to develop digital assistant is easily extensible to other areas and data fields.
    \item Verdict Accuracy: While not the primary focus of this evaluation, Reasoning Engine showed approximately 93.33\% accuracy in determining scam verdicts, although this result is based on a small sample size.
\end{itemize}

Manual evaluation was chosen over automated evaluation to maintain the quality and depth of the assessment. This also enables the experts to tease out small nuances in features, and if the model is correctly able to identify these. The results demonstrate digital assistant's potential to significantly improve the efficiency and accuracy of scam reviews, with the added benefit of uncovering new reasoning missed by human reviewers.

The expert evaluation revealed that the digital assistant significantly improved the efficiency and accuracy of scam reviews. Reviewers reported that the explanations provided by the digital assistant were clear, concise, and helpful in understanding the underlying reasons for a scam classification. The assistant also enabled reviewers to process a larger volume of transactions in less time, thereby increasing their overall productivity.

%% file: generalization_guidelines.tex
\section{Generalization to Other Transaction Platforms}
\label{sec:generalization_to_other_upi_platforms}
This research focused on GPay as a use case, but the developed approach is inherently generalizable to other UPI-enabled payment platforms and even beyond the payments domain. The LLM-based classifier and the reasoning engine can be adapted to different platforms by training them on platform-specific data and incorporating relevant features. The flexibility of LLMs allows for the customization of prompts and training data to cater to the specific nuances and characteristics of each platform. For instance, the LLM-based classifier and reasoning engine could be adapted to identify and prevent online harassment, hate speech, misinformation, and other forms of harmful content.  The ability of LLMs to understand and interpret natural language makes them a powerful tool for addressing a wide range of Trust \& Safety challenges in the digital age.

%% file: limitations.tex
\section{Limitations} \label{sec:limitations}
While the LLM-based approach presented in this paper shows promising results in scam detection, it is important to acknowledge its limitations. Firstly, the model may need be constantly updated with latest types of scams or scammer's evolving tactics, it is important to see how does this model hold over time. Secondly, the interpretability of the LLM's decisions can be a challenge, as the model's reasoning may not always be transparent or easily understandable by human reviewers. Thirdly, the model's performance may vary across different transaction types and platforms, requiring careful customization and fine-tuning for optimal results. Finally, the computational cost associated with LLM inference can be a significant factor, especially when dealing with large volumes of transactions.

Future research should focus on addressing these limitations by exploring ways to enhance the model's interpretability, improving its performance on diverse transaction types, and optimizing its computational efficiency.

%% file: real_world_impact.tex
\section{Real-World Impact and Future Work}\label{lm:real_world_impact}

Fraud \& scam has a real economic and human cost. Consumers in the US lost 10 billion dollars to fraud in 2023 \cite{kapron2023google, gupta2024online} (with investment scams being the leading category), and in India (where this research got its start with GPay), digital payments scam India grew five-fold in 2024 \cite{gupta2024online} from one year ago.

The successful deployment of LLM-based scam detection systems has the potential to significantly reduce financial losses due to scams in digital payment systems. By automating the identification of suspicious transactions and providing human reviewers with actionable insights, the system can enhance the efficiency and accuracy of scam detection and reviews, leading to faster, more accurate, and more effective mitigation of fraudulent activities.

On the technical front, future work can be undertaken in several directions, three of which are on-device models, distillation, and RLHF. First, future models will likely be small enough to deploy directly on a payment device such as a mobile phone, giving relatively quick response. Second, distillation can help reduce the feature space and model size to have similar effectiveness with a much smaller and condensed parameter space, thus driving efficiency. Third, RLHF mechanisms can be built into the digital assistant so that if it gets an evaluation wrong, expert reviewers can train it to do better (a 'thumbs up' / 'thumbs down' followed by a feedback field with explanation on how to evaluate a particular case).

Beyond payments, scam detection as described in our work can be extended to several other domains. E-commerce, B2B payments, 3rd party merchant transactions, invoicing, tax scams, and other such areas also get targeted by scammers quite frequently. So long as model features and prompts are appropriately designed, guidelines (based on known modus operandi) are fed to the LLM, and sufficient controls are established to ensure validity of results through automated or human evaluations (evals), the approach is extensible to a vast array of digital scam prevention.

Many LLM models are inherently internationalized because of in-built support for many languages. Because most LLMs now support international languages, this also presents an opportunity to extend a set of learnings from one geography to others with simple prompting. Moreover, simple tweaks can be made to prompts based on local context (in a human readable, interpretable form) that can help extend a base model to a localized "expert" scam detector.

Additionally, it is conceivable to extend this approach beyond digital and text-based scam to other modalities. For example, a phone call can be transcribed and run through a scam detector (in real-time). Our 'classifier and assistant' approach, based on human-provided guidelines and up-to-date refinements is \textit{generally} extensible. It can deal relatively easily with the latest type of scam actions and modalities with simple, centralized prompt updates, reducing or preventing the delay, cost, and effort involved in disseminating new guidelines to hundreds or thousands of (geographically distributed) reviewers through training and documentation. With these operational efficiencies, reviewers can focus on high value, high importance cases and rely on the model to stay up-to-date on guidelines \& policies and give them the insights they need to do their jobs well.

The research highlights the potential of LLMs in addressing various Trust \& Safety challenges, paving the way for further exploration of these powerful models in promoting a safer and more secure digital environment.

%% file: conclusion.tex
\section{Conclusion} \label{sec:conclusion}
In conclusion, this paper demonstrates the significant potential of Large Language Models (LLMs) in revolutionizing scam detection within digital payment systems. By harnessing the power of LLMs to analyze textual data and discern intricate patterns, we have developed a robust and adaptable scam detection solution that can be applied to various platforms.

Our findings reveal that LLMs, when fine-tuned on labeled datasets and guided by carefully crafted prompts, can achieve high recall rates in identifying scam transactions while maintaining reasonable precision. This is particularly noteworthy in high-value segments and transactions with descriptive text, where traditional ML models often struggle. Moreover, the development of the digital assistant, powered by an LLM-based reasoning engine, has proven to be a valuable tool for human reviewers, enhancing their efficiency and accuracy in the scam review process.

While this work primarily focused on GPay, the methodology and framework presented here are inherently generalizable to other UPI-enabled platforms and even extendable to other domains within Trust \& Safety. The flexibility of LLMs allows for seamless adaptation to different platforms and use cases, making them a powerful tool for addressing a wide range of challenges in the digital age.

However, it is important to acknowledge the limitations of the LLM-based approach detailed in the section \ref{sec:limitations}. Future research should focus on addressing these limitations by exploring ways to enhance the model's interpretability, improving its performance on diverse transaction types, and optimizing its computational efficiency.

Despite these challenges, the potential of LLMs to transform the financial fraud detection landscape is undeniable. By augmenting existing ML models and empowering human reviewers, LLMs can significantly contribute to creating a safer and more secure digital payment ecosystem. As research in this field progresses, we can anticipate even more sophisticated and effective LLM-based solutions that will further strengthen the fight against scams and fraud in the digital age.

%% file: future_scope.tex
\section{Future Scope} \label{sec:future_scope}
\begin{itemize}
    \item {Bias Monitoring and Mitigation:} We will continuously monitor the scam classification and reasoning system for potential biases that may arise during its operation. If any biases are identified, we will retrain the models on carefully curated datasets to counteract these biases and ensure fairness and accuracy in scam detection and reasoning.
    \item {Accuracy and Efficiency Enhancements:} We will continue to refine our scam classification and reasoning system by leveraging the latest advancements in Gemini Models.
\end{itemize}